\documentclass[12pt]{article}
\usepackage{graphicx}

\newcommand {\be}{\begin{equation}} 
\newcommand{\ee}{\end{equation}}    

\def\dds1{\frac{\partial}{\partial s_1}}

\def\d{d\kern-0.8 ex\vrule height 1.3 ex depth-1.24 ex width 0.7 ex
\kern 0.15 ex}
\def\D{D\kern-1.7 ex\vrule height .87 ex depth-0.8 ex width 0.7 ex
\kern 0.95 ex}

\begin{document}
\baselineskip 20 pt

\begin{center}

\Large{\bf Acceleration of dust particles by vortex ring}

\end{center}

\vspace{0.5cm}

\begin{center}
{\bf  Zahida Ehsan$^{a,b}$, N. L. Tsintsadze$^{a,c}$,  J. Vranjes$^d$,  M. Coppins$^b$,\\
 R. Khan$^{e}$, S. Poedts$^d$, and J. E.
Allen$^f$ }

\end{center}

\begin{center}

$^a$Salam Chair in Physics, G.C. University, Lahore, 54000, Pakistan.

\vspace{0.3cm}

$^b$Plasma Physics Group, Blackett Laboratory, Imperial College London,\\
SW72AZ, United Kingdom.

\vspace{0.3cm}

$^c$Department of Plasma Physics, E. Andronikashvili Institute of Physics,\\ 0171 Georgia.

\vspace{0.3cm}

$^d$Center for Plasma Astrophysics, and Leuven Mathematical Modeling \\ and Computational Science Centre
 (LMCC),  Celestijnenlaan 200B, \\ 3001 Leuven,  Belgium.

\vspace{0.3cm}

$^e$National Tokamak Fusion Program, PO Box 3329, PAEC Islamabad, Pakistan

\vspace{0.3cm}

$^f$University College, Oxford OX1 4BH, and OCIAM, \\Mathematical Institute, Oxford OX1 3LB, UK

\end{center}

\vspace{1cm}

{\bf Abstract:}  It is shown that nonlinear interaction between large amplitude circularly polarized EM wave and dusty
plasma leads to a nonstationary ponderomotive force which in turn produces a vortex ring, and magnetic field. Then the
ensuing vortex ring in the direction of propagation of the pump wave can accelerate the micron-size dust particles
which are initially at rest and eventually form a non relativistic dust jet. This effect is purely nonstationary and
unlike linear vortices, dust particles do not rotate here. Specifically, it is pointed out that the vortex ring or
closed filament can become potential candidate for the acceleration of dust in tokamak plasmas.

\vspace{1cm}


\pagebreak


In the past there have been fairly extensive investigations in the field of nonlinear interaction of high frequency EM
waves or short pulse laser beams with an electron-ion plasma \cite {1,2,3,4,5,6,7,8,9,10,11,11',12,13',13'',13,14,15}.
Such interactions can produce various types of phenomena, like self focusing, Brilliouin or Raman scattering,
filamentation or modulational instability, collisionless shock waves or solitons, wave breaking, absorption of EM
waves, generation of vortex rings, quasi-static magnetic fields etc. Spontaneously generated magnetic fields are not
only important for laboratory produced plasmas, but in many cosmic environments, in our Universe as well as in galactic
and intergalactic spaces\cite{17,18,19,20,21,22,23}. A mechanism for the generation of large quasi-static magnetic
fields and vortex ring by a non-potential ponderomotive force resulting from the time dependent amplitude of EM waves
in electron ion plasma was discussed in Ref. \cite{5}. This way production of magnetic field can essentially affects
the transport phenomena, as well as the absorption of an EM field in inertial confinement fusion (ICF) schemes. Indeed,
the vortex ring phenomena can have profound influence on laser experimental research, and at the same time its behavior
can be described within the main equations of a continuous medium \cite {12,13',13''}. Most significantly, once this
vortex is generated, it develops only under the effect of its own dynamics. Unlike straight vortex filament that is a
static phenomenon of plasma, the vortex ring moves relative to the plasma, and in the same time it expands noticeably
in the process.

Recently, research on dust in tokamak plasmas has attracted a lot of interest, although existence of these micron-size
particles has been known for a long time.\ Besides some safety threats and engineering issues, the hereness of these
such heavier particles in fusion devices can influence the plasma operation and performance effectively. Thus it is
expected that dust in plasmas is going to play requisite/essential role in the next generation ITER-like fusion
devices\cite{27,29,29',30',30,31,32}.

It has been predicted theoretically that dust particles in tokamak plasma can gain very high velocity and their
presence can significantly affects when these collide the tokamak wall, the ejecta far exceed the projectile masses.
This not only supplies fresh particles but releases neutral gas and a runaway effect for wall erosion and plasma
contamination that could be a potential hazard for maintaining fusion conditions\cite{30}. Also if such fast particles
exist in the scrape-off layer (SOL), dust impact ionization could be used as a diagnostic. Different mechanisms have
been proposed which can cause the acceleration of dust to such velocities. Among them, ion/dust drag force is mostly
believed to play dominant role the values of which strongly depend on plasma parameters as well as on dust size, shape,
temperature, and electric charge, it could speed up the dust up to ion flow velocities, which are found to be tens of
km/s for toroidal plasma flow near the last closed magnetic surface (LCMS)\cite{30}.

deAnglis et al proposed a mechanism for the acceleration of dust based on stochastic heating\cite{31}. Shukla and
Tsintsadze quite recently showed that the normal component of the space charge electric field may accelerate dust
particles in the scrape-off layer close to the tokamak chamber wall\cite {32}.

In this article we propose a mechanism for the acceleration of dust particles by closed filaments or vortex rings,
which are produced by the action of nonstationary ponderomotive force occurring due to the time dependence of the
amplitude of EM wave, which may have effects on the physics of tokamak plasmas.

Following standard technique for the interaction of a high-frequency electromagnetic radiation with some accidental and
relatively low-frequency plasma perturbations, we write down the equation of motion for dust grains and ions{\bf ,}
\begin{equation}
\frac{\partial {\bf v}_{d}}{\partial t}+({\bf v}_{d}\cdot \nabla ){\bf v}%
_{d}=-\frac{Z_{d}e}{m_{d}}\left( {\bf E}+\frac{1}{c}{\bf v}_{d}\times {\bf B}%
\right) ,  \label{1}
\end{equation}
\begin{equation}
m_{i}\left[ \frac{\partial {\bf v}_{i}}{\partial t}+({\bf v}_{i}\cdot \nabla
){\bf v}_{i}\right] =e\left( {\bf E}+\frac{1}{c}{\bf v}_{i}\times {\bf B}%
\right) +{\bf F}_{NL}-\frac{1}{n_{i}}\nabla P_{i}.  \label{2}
\end{equation}
The coupling with the intense EM field is described by the time dependent ponderomotive force \cite{4,6,13,34}
expressed in terms of the vector potential
\[
{\bf F}_{NL}=-m_{i}c^{2}\left[ \frac{\omega _{o}}{\omega _{o}-\omega _{ci}}%
\nabla \left( \frac{e^{2}|A^{2}|}{m_{i}^{2}c^{4}}\right) -\frac{\omega _{ci}%
{\bf k}_{o}}{\left( \omega _{o}-\omega _{ci}\right) ^{2}}\frac{\partial }{%
\partial t}\left( \frac{e^{2}|A^{2}|}{m_{i}^{2}c^{4}}\right) \right] .
\]
It is to be noted that above ponderomotive force expression contains two parts, space dependant and time dependant,
where the later is due to magnetic field. In the absence of static magnetic field ($B_{0}=0$) the ponderomotive force
expression reduces to the ordinary one\cite{35}.In an electron-depleted plasma, or equivalently for oscillations well
below the electron Debye scale, the ions play a vital role for the excitation of a very low frequency wave. Hence, the
quasi-neutrality condition which we use reads $\delta n_{i}\simeq Z_{d}\delta n_{d}${\bf . }Neglecting the inertial
term in Eq. (\ref{2}), and combining the remaining equation with Eq. (\ref{1}%
) in the limit $v_{d},v_{i}<<c,$ we get
\begin{equation}
\frac{\partial {\bf v}_{d}}{\partial t}=-\frac{1}{2m_{d}}\nabla \frac{%
v_{d}^{2}}{2}+{\bf v_{d}}\times (\nabla \times {\bf v}_{d})-\frac{%
Z_{d}m_{i}c^{2}}{m_{d}}\left[ \frac{\omega _{o}}{\omega _{o}-\omega _{ci}}%
\nabla \Psi ^{2}-\frac{\omega _{ci}{\bf k}_{o}}{\left( \omega _{o}-\omega
_{ci}\right) ^{2}}\frac{\partial \Psi ^{2}}{\partial t}\right] -\frac{Z_{d}}{%
n_{i}m_{d}}\nabla P_{i},  \label{3}
\end{equation}
where $\Psi =eA/(m_{i}c^{2})$ is a dimensionless vector potential. Taking the curl of both sides, we obtain
\begin{equation}
\frac{\partial {\bf \Omega }}{\partial t}=\nabla \times {\bf v}_{d}\times {\bf \Omega }+\frac{Z_{d}\omega
_{ci}m_{i}c^{2}}{m_{d}\left( \omega
_{o}-\omega _{ci}\right) ^{2}}{\bf \nabla }\frac{\partial \Psi ^{2}}{%
\partial t}\times {\bf k}_{o}.  \label{4}
\end{equation}
here ${\bf \Omega }=\nabla \times {\bf v}_{d}$ represents the vorticity of the dust velocity, and ${\bf k}_{o}$ is the
wave vector which is directed along the external magnetic field [${\bf k}_{o}=(0,0,k_{o}{\bf e}_{z})]$. Considering
$l>v_{d}t$, where $l$ and $t$ represent the characteristic spatial and time scale lengths, respectively, will allow us
to neglect the first term on the right-hand side (r.h.s) in comparison with the term on the left-hand side (l.h.s) in
Eq. (\ref{4}). As a result, we obtain a simple relation between the vorticity and the source
\begin{equation}
{\bf \Omega }=\frac{Z_{d}\omega _{ci}m_{i}c^{2}}{m_{d}\left( \omega _{o}-\omega _{ci}\right) ^{2}}{\bf \nabla }\Psi
^{2}\times {\bf k}_{o}. \label{5}
\end{equation}
For a circularly polarized EM wave propagating along the $z$-axis, the
vorticity has only two components, ${\bf \Omega }_{x}$ and ${\bf \Omega }%
_{y} $. Thus vortices are produced in the plane perpendicular to the propagation of the pump wave, i.e., we have the
formation of vortex rings or close filaments.

Now we investigate Eq. (\ref{4}) in the cylindrical coordinates $r,\theta ,z$%
\ at the center of the ring, which clearly shows that the vorticity has only one component, ${\bf \Omega =\Omega
}_{\theta }{\bf e}_{\theta }$, and the
particle velocity is ${\bf v}_{d}=v_{r}\cdot {\bf e}_{r}+v_{z}\cdot {\bf e}%
_{z}$, where $({\bf e}_{r},{\bf e}_{\theta },{\bf e}_{z})$\ are the unit
vectors{\bf . }Expressing{\bf \ }$({\bf \Omega \cdot \nabla })=\left( {\bf %
\Omega }_{\theta }/r\right) \left( \partial /\partial \theta \right) ${\bf ,
}$\partial v/\partial \theta =v_{r}{\bf e}_{\theta }${\bf , }and{\bf \ }$%
v_{r}=dr/dt${\bf \ }in Eq. (\ref{4}), we obtain
\begin{equation}
\frac{d}{dt}\left( \frac{\Omega _{\theta }}{n_{d}r}\right) =\frac{%
Z_{d}\omega _{ci}m_{i}c^{2}}{m_{d}\left( \omega _{o}-\omega _{ci}\right) ^{2}%
}\frac{k_{o}}{n_{d}r}\frac{\partial }{\partial r}\left( \frac{\partial \Psi ^{2}}{\partial t}\right) .  \label{6}
\end{equation}
From Eqs. (\ref{3})-(\ref{6}) it can be noticed that the time dependence of the ponderomotive force does not conserve
the velocity circulation, $\Gamma _{d}=\oint $ ${\bf v}_{d}{\bf dr=}\int_{s}\int {\bf \Omega ds}$. The same follows for
frozen-in condition\cite{36} which does not take place but only
for the wake vorticity ($\partial \Psi ^{2}/\partial t=0)$%
\begin{equation}
\frac{\Omega _{\theta }}{n_{d}r}=const.  \label{7}
\end{equation}
Eqs. (\ref{4})-(\ref{6}) indicate that if at the initial instant of time there are no vortices of the dust fluid flow
at a given point, they will be generated by the circularly polarized EM field. It is important to note that if there
are $N$ point-like vortices (filaments), then these filaments can interact with each other, leading eventually to the
merging of vortices, to the decaying of a vortex into two other vortices, to the annihilation of vortices, etc.{\bf \
}Another interesting feature is that here unlike linear vortices, dust particles do not rotate.

We have shown above how a circularly polarized EM wave generates closed filaments, values of which can be defined by
the intensity of the pump. We now demonstrate that there is a mechanism of the acceleration of the dust particles up to
a particular velocity due to the vortex ring. For this considering the inverse problem, i.e., by a given vortex we can
define the velocity of the dust grains (${\bf v}_{d}$) at any point in the plasma. To this end, we assume that the
plasma is at rest until the vortex is generated, ${\bf \Omega }={\bf \nabla \times v_{d}}$. Let us introduce the vector
${\bf P}$\ such that $\nabla \cdot {\bf P}=0$, and ${\bf v_{d}=\nabla \times P}$. For ${\bf P}$\ we can write down the
following equation
\begin{equation}
{\bf \nabla }^{2}{\bf P}=-{\bf \Omega },  \label{8}
\end{equation}
which has solution of the form
\begin{equation}
{\bf P}=\frac{1}{4\pi }\int \frac{{\bf \Omega }({\bf r}^{^{\prime }})d{\bf r}%
^{^{\prime }}}{R}.  \label{9}
\end{equation}
where $d{\bf r}^{\prime }=dx^{\prime }dy^{\prime }dz^{\prime }$, $R=\left[ (x-x^{\prime })^{2}+(y-y^{\prime })^{2}+(\xi
-z^{\prime })^{2}\right] ^{1/2}, $ $\xi =\zeta -u_{g}t$ and{\bf \ }$u_{g}=\frac{2k_{o}c^{2}\omega _{cd}}{\omega
_{pd}^{2}}${\bf \ }is the group velocity of the pump wave.

By knowing the vector ${\bf P}$, we can define the velocity ${\bf v}_{d}$ at any point in the plasma
\begin{equation}
{\bf v}_{d}({\bf r},t)={\bf \nabla }\times \frac{1}{4\pi }\int \frac{\Omega (%
{\bf r}^{\prime })d{\bf r}^{\prime }}{R}.  \label{10}
\end{equation}
Let us now consider a vortex ring with radius $\rho _{o}$, and characterize the position of the variable on the vortex
ring by the angle $\alpha $. Any
point on the vortex filament in the Cartesian coordinates is determined by $%
x^{\prime }=\rho _{o}\cos \alpha $, $y^{\prime }=\rho _{o}\sin \alpha ,$\ $%
z^{\prime }=0$.

In order to calculate the integral in Eq. (\ref{9}), we assume that $\Psi ^{2}=\Psi _{0}^{2}u_{g}\tau \delta (z^{\prime
}-u_{g}\tau )\theta (r^{\prime }-\rho _{o})$, and ${\bf P}$\ has only one component, $P_{\theta }=P_{\theta }({\bf
r},z)$\cite{36}$.$ From Eq. (\ref{9}), we have
\begin{equation}
P_{\theta }(r,\xi )=\frac{\omega _{ci}k_{o}u_{g}\tau \rho _{o}\Psi ^{2}}{%
m_{d}\pi \left( \omega _{o}-\omega _{ci}\right) ^{2}}\oint \frac{d\alpha \cdot \cos \alpha }{R},  \label{11}
\end{equation}
here $R=(r^{2}+\xi ^{2}+\rho _{o}^{2}-2r\rho _{o}\cos \alpha )^{1/2}$, $\tau ${\bf \ }is the duration of EM field and
the integral is
\begin{equation}
\oint \frac{\cos \alpha d\alpha }{R}=\frac{4}{\eta }\sqrt{\frac{1}{r\rho _{o}%
}}\left[ \left( 1-\frac{\eta ^{2}}{2}\right) K(\eta )-E(\eta )\right] . \label{12}
\end{equation}
where $\eta ^{2}=4r\rho _{o}/[(r+\rho _{o})^{2}+\zeta ^{2}]$, $K$ and $E$ are the elliptical integrals of the first and
second kind,
\[
K(\eta )=\int_{0}^{2\pi }\frac{d\beta }{\sqrt{1-\eta ^{2}\sin ^{2}\beta }},
\]
\[
E(\eta )=\int_{0}^{2\pi }\sqrt{1-\eta ^{2}\sin ^{2}\beta }d\beta ,
\]
and $\beta =(\alpha -\pi )/2$. As a result we have
\begin{equation}
P_{\theta }(r,\xi )=\frac{\omega _{ci}k_{o}u_{g}\tau \Psi _{0}^{2}}{%
m_{d}\left( \omega _{o}-\omega _{ci}\right) ^{2}\eta }\times \sqrt{\frac{%
\rho _{o}}{r}}\left[ \left( 1-\frac{\eta ^{2}}{2}\right) K(\eta )-E(\eta )\right] .  \label{13}
\end{equation}
The components of the velocity can be written as
\begin{equation}
v_{dr}=-\frac{\partial P_{\theta }}{\partial z},\quad v_{dz}=\frac{1}{r}%
\frac{\partial }{\partial r}(rP_{\theta }),  \label{14}
\end{equation}
or
\begin{equation}
v_{dr}=\frac{\omega _{ci}k_{o}u_{g}\tau \Psi _{0}^{2}}{m_{d}\left( \omega
_{o}-\omega _{ci}\right) ^{2}2\pi r\sqrt{(r+\rho _{o})^{2}+\zeta ^{2}}}%
\times \zeta \left[ -K(\eta )+\frac{r^{2}+\rho _{o}^{2}+\zeta ^{2}}{(\rho _{o}-r)^{2}+\zeta ^{2}}E(\eta )\right] ,
\label{16}
\end{equation}
\begin{equation}
v_{dz}=\frac{\omega _{ci}k_{o}u_{g}\tau \Psi _{0}^{2}}{m_{d}\left( \omega _{o}-\omega _{ci}\right) ^{2}2\pi
\sqrt{(r+\rho _{o})^{2}+\zeta ^{2}}}\times \left[ K(\eta )+\frac{r^{2}+\rho _{o}^{2}+\zeta ^{2}}{(\rho
_{o}-r)^{2}+\zeta ^{2}}E(\eta )\right] .  \label{17}
\end{equation}
Now we consider two cases, for the first, assuming $r\rightarrow 0$ i.e., on the axis
\begin{equation}
v_{dr}=0,\qquad v_{dz}=\frac{\omega _{ci}k_{o}u_{g}\tau \Psi _{0}^{2}}{%
m_{d}\left( \omega _{o}-\omega _{ci}\right) ^{2}\left( \rho _{o}^{2}+\zeta ^{2}\right) ^{3/2}}.  \label{18}
\end{equation}
Recalling that $\xi -\zeta -u_{g}t=0$, and for the maximum, we write
\begin{equation}
v_{dz,\max }=\frac{\omega _{ci}k_{o}u_{g}\tau \Psi _{0}^{2}}{m_{d}\rho _{o}\left( \omega _{o}-\omega _{ci}\right)
^{2}}.  \label{19}
\end{equation}
Now we consider the case at the points near the filament, i.e., $r=\rho _{o}$
and $\xi =\zeta -u_{g}\tau =0.$ In this case $\eta ^{2}=1,$ $E=\pi /2$, but $%
K(1)$ is logarithmically divergent at the lower limit as
\begin{equation}
K=\frac{1}{2}\int_{0}^{\pi }\frac{d\alpha }{\sin \alpha /2}\simeq \int_{0}^{\pi }\frac{d\alpha }{\alpha }.  \label{20}
\end{equation}
In reality, the closed filament as a ring has a finite size denoted by $%
r_{o} $ $(r_{o}$ is the core radius of the thin vortex ring), so there must be a cut off at a value $\alpha \sim
r_{o}/\rho _{o}$, and $K(1)=\ln \rho _{o}/r_{o}$. The $z$-component of the velocity now becomes
\begin{equation}
v_{dz}=\frac{\omega _{ci}k_{o}u_{g}\tau \Psi _{0}^{2}}{m_{d}\rho _{o}\left( \omega _{o}-\omega _{ci}\right)
^{2}}ln\frac{\rho _{o}}{r_{o}}.  \label{21}
\end{equation}
Using the simple relation between vorticity and source we have obtained expressions of dust speed [Eqs.(12) and (13)]
which particles gain from the filaments and have shown vortex ring can generate a collimated dust jet along the
propagation of EM wave. We emphasis that above consideration can be applied to the acceleration of dust particles in
the scrape-off layer of the Tokamak.

Now we manifest how the nonstationary ponderomotive force of the EM wave which creates slowly varying electric fields
and vector potentials can generate magnetic fields.

A simple expression for this can be obtained by taking the curl of the momentum equation for inertialess ions
\begin{equation}
\nabla \times {\bf E}=-\frac{1}{e}\nabla \times {\bf F}_{NL}.  \label{22}
\end{equation}
Using the Faraday law $\nabla \times {\bf E}=-(\partial {\bf B}/\partial t)/c $, Eq. (\ref{22}) becomes
\begin{equation}
\frac{\partial {\bf B}}{\partial t}=\frac{c}{e}\nabla \times {\bf F}_{NL}=%
\frac{c}{e}\left[ \frac{\omega _{ci}m_{i}c^{2}}{\left( \omega _{o}-\omega _{ci}\right) ^{2}}{\bf \nabla }\frac{\partial
\Psi ^{2}}{\partial t}\times {\bf k}_{o}\right] .  \label{24}
\end{equation}
Integrating both sides we get
\begin{equation}
{\bf B}=\frac{\omega _{ci}m_{i}c^{3}}{e\left( \omega _{o}-\omega _{ci}\right) ^{2}}{\bf \nabla }\Psi ^{2}\times {\bf
k}_{o}.  \label{25}
\end{equation}
The $x$ and $y$ components of the above equation are:
\begin{equation}
\frac{eB_{x}}{m_{i}c}=\frac{c^{2}\omega _{ci}{\bf k}_{o}}{\left( \omega _{o}-\omega _{ci}\right) ^{2}}\frac{\partial
\Psi ^{2}}{\partial y}, \label{26}
\end{equation}
\begin{equation}
\frac{eB_{y}}{m_{i}c}=-\frac{c^{2}\omega _{ci}{\bf k}_{o}}{\left( \omega _{o}-\omega _{ci}\right) ^{2}}\frac{\partial
\Psi ^{2}}{\partial x}. \label{27}
\end{equation}
Squaring and then addition of Eqs. (\ref{26}) and (\ref{27}) gives us generation of magnetic field
\begin{equation}
\frac{eB}{m_{i}c}=\pm \frac{k_{o}\omega _{ci}}{m_{i}(\omega _{o}-\omega _{ci})^{2}}\frac{\partial \Psi ^{2}}{\partial
r},\quad r=(x^{2}+y^{2})^{1/2}. \label{28}
\end{equation}
Eqs. (\ref{26})-(\ref{28}) show that magnetic field is generated in the plane perpendicular to the starting magnetic
field $B_{o}{\bf e}_{z}$, and the components of the generated magnetic field are determined by both, the parameters of
the plasma, and the intensity of the pump wave.

To summarize, the main idea here is to identify a new mechanism for the generation of vortex ring which can accelerate
dust particles to very high velocity resulting in the formation of a non relativistic dust jet by employing the
nonlinear interaction of circularly polarized EM wave with dusty plasma. Specifically, this interaction leads to a
nonstationary ponderomotive force which pushes the ions locally and creates slowly varying electric fields and vector
potentials. The latter, in turn, generates vortex ring and quasi stationary magnetic fields in the perpendicular
direction. Then, we considered an interesting inverse problem that the ensuing vortex ring becomes a means for the
formation of dust jets (dust particles were initially in an equilibrium state), which are emitted out after gaining
acceleration. The relevance of the above analysis for tokamak plasmas can be assessed by making some simple estimates.
In addition, generation of magnetic field is a ubiquitous phenomenon relevant to astrophysics and this can also affect
energy transport in inertial confinement fusion schemes.


\begin{thebibliography}{99}
\bibitem{1}  Akhiezer, I. A., et al. 1964 Sov. Phys. JETP {\bf 19}, 208.

\bibitem{2}  Zakharov, V 1972 Sov. Phys. JETP {\bf 35}, 908.

\bibitem{3}  Tsintsadze, N.L. 1974 Phys. Lett. {\bf 50A}, 33.

\bibitem{4}  Karpman, V. I. and Washimi, H. 1976 Sov. Phys. JETP 44, 528.

\bibitem{5}  Berezhiani, V. I., et al. 1980 J. Plasma. Phys. {\bf 24}, 15.

\bibitem{6}  Tsintasdze, N. L. and Watanabe, M. 1980 Sov. J. Plasma Phys.%
{\bf \ 6, }6.

\bibitem{7}  Tsintsadze, N. L., et al 2002 Phys. Plasmas{\bf . 9 }4270.

\bibitem{8}  Freund, H. P., et al 1981 J. Plasma Phys. {\bf 25}, 465.

\bibitem{9}  Stamper, J. A. 1992 Las. Part. Beams {\bf 9}, 841.

\bibitem{10}  Tsintsadze, L. N., Callebaut, D. K. and Tsintsadze, N. L. 1996
J. Plasma Phys. {\bf 55}, 407.

\bibitem{11}  Sudan, R. N. 1993 Phys. Rev. Lett. {\bf 70}, 3075; Askaryan,
G. A.,Bulanov, S. V., Pegoraro, F., and Pukhov, A. M., 1994 Sov. Phys. JETP {\bf 60}, 251 1994; Gorbunov, L. M., Mora,
P., and Antonsen, T. M., 1996 Phys. Rev. Lett. {\bf 76}, 2945.

\bibitem{11'}  Borghesi, M., et al. 1998 Phys. Rev. Lett. {\bf 81}, 112.

\bibitem{12}  Wilks, S. C., et al., 1992 Phys. Rev. Lett. 69, 1383.

\bibitem{13'}  Tsintsadze, L. N., Mima, K. andNishikawa, K. 1998 Plasma
Phys. Controlled Fusion 40, 1933.

\bibitem{13''}  Tsintsadze, L. N., Nishikawa, K.,Tajima, T., and Mendonca,
J. T. 1999 Phys. Rev. E {\bf 60}, 7435.

\bibitem{13}  Tsintsadze, L. N, Pajouh, H. H, Tsintsadze, N. L. ,Mendonca,
J. T, and Shukla, P. K., 2000 Phys. Plasmas{\bf \ 7, }2348.

\bibitem{14}  Vranjes, J. ,Saleem, H, andPoedts, S., 2007 Phys. Plasmas {\bf %
14}, 034504.

\bibitem{15}  Nakamura,T and Mima, K., 2008 Phys. Rev. Lett. {\bf 100},
205006.

\bibitem{17}  Tatarakis, M. et al., 2002 Nature London{\bf \ 415},
280;Wagner, U. et al., 2004 Phys. Rev. E {\bf 70}, 026401.

\bibitem{18}  Kolb E. W., and Turner, M. S. 1994. The Early Universe
Addison-Wesley, Reading.

\bibitem{19}  Ryu, D. et al., 2008 Science{\bf \ 320}, 909 (2008).

\bibitem{20}  Widrow, L. M. 2002 Rev. Mod. Phys. {\bf 74}, 775.

\bibitem{21}  Kulsrud, R. M., and Zweibel, E. G. 2008 Rep. Prog. Phys. {\bf %
71}, 046901.

\bibitem{22}  Bernet, M. L. et al., 2008 Nature (London) {\bf 454}, 302.

\bibitem{23}  Bobin, J. L., 1983 Phys. Rep. {\bf 122}, 173.

\bibitem{24}  Hatchett, S. P., et al., 2000 Phys. Plasmas {\bf 7}, 2076.

\bibitem{25}  Snavely, R. A., et al., 2000 Phys. Rev. Lett. {\bf 85}, 2945.

\bibitem{26}  Wilks, S. C., et al., 1992 Phys. Rev. Lett. {\bf 69}, 1383.

\bibitem{27}  Narihara, K. , 2001 Nucl. Fusion {\bf 41} 1967.

\bibitem{29}  Rudakov, D. L. et al., 2007 J. Nucl. Matter. {\bf 227}, 363.

\bibitem{29'}  Castaldo, C. et al., 2007 Nucl Fusion{\bf \ 47}, L5-L9.

\bibitem{30'}  Ratynskaia,\ S., 2008 Nucl Fusion {\bf 48}, 015006.

\bibitem{30}  de Angelis, U., Marmolino, C. , and Tsytovich, V. 2005 Phys.
Rev. Letters, {\bf 95}, 095003.

\bibitem{31}  de Angelis, U., Ivlev,A. ,Tsytovich, V. and Morfill, G. 2005
Phys. Plasmas, {\bf 12}, 052301.

\bibitem{32}  Shukla, P. K. and Tsintsadze, N. L., 2008 Phys. Lett. A {\bf %
372}, 2053.

\bibitem{33}  Ehsan, Z. , Tsintsadze, N. L. , Murtaza, G. and Shah, H. A.
2009 Phys. Plasmas {\bf 16}, 023702.

\bibitem{34}  Ehsan, Z. , Tsintsadze, L. N. Vranjes, J. and Poedts, S. 2009
Phys. Plasmas{\bf \ 16, }053702.

\bibitem{35}  Tsintsadze, N. L. , Ehsan, Z, Shah, H. A. , and Murtaza, G.,
2006 Phys. Plasmas{\bf 13}, 072103; Proceedings of the Second International Symposium on Unconventional Plasmas,
Eindhoven, Netherlands, 2006, edited by Callebaut D. K, EUT, Eindhoven, 2006 , Vol. 161.

\bibitem{36}  Saffman, P. G. , 1995 {\em Vortex Dynamics} (Cambridge
University Press, Cambridge).
\end{thebibliography}
\end{document}